\documentstyle[sprocl,epsfig]{article}

\def\NPB{{\em Nucl. Phys.} B}
\def\PLB{{\em Phys. Lett.}  B}
\def\PRL{\em Phys. Rev. Lett.}
\def\PRD{{\em Phys. Rev.} D}

\def\be{\begin{equation}}
\def\ee{\end{equation}}
\def\bea{\begin{eqnarray}}
\def\eea{\end{eqnarray}}

\begin{document}

\rightline{KEK-TH-988}
\rightline{OU-HET 491}
\vspace{1cm}
\title{NEW PHYSICS FROM HIGGS SELF-COUPLING MEASUREMENT
\footnote[1]{Talk given at International Conference on Linear Colliders (LCWS 2004), April 19-23,
 2004, Paris, France.}\\}
\author{ SHINYA KANEMURA$^{a}$, YASUHIRO OKADA$^{b,c}$, 
and EIBUN SENAHA$^{b,c}$ }

\address{ $^a$Department of Physics, Osaka University, Toyonaka, Osaka
 560-0043, Japan\\
$^b$Theory Group, KEK, Tsukuba, Ibaraki 305-0801, Japan\\
$^c$Department of Particle and Nuclear Physics, 
         the Graduate University for Advanced Studies (SOKENDAI), 
         Tsukuba, Ibaraki 305-0801, Japan}


\maketitle\abstracts{
Radiative correction to the triple Higgs boson coupling is studied
in the two Higgs doublet model in connection with the electroweak 
baryogenesis. It is shown that the one loop correction
is large enough to be identified in future linear collider experiments, 
if the baryon number of the Universe is generated
at the first order electroweak phase transition. 
}
  
\section{Introduction}
Measurement of the triple Higgs boson coupling is one of the most 
important goals of the Higgs physics at a future $e^+e^-$ linear collider
(LC) experiment. This would provide the first direct information on the Higgs 
potential that is responsible for electroweak symmetry breaking.
The existence of the triple Higgs coupling is a characteristic feature of
the spontaneous breaking of the electroweak symmetry in models with only 
weak-doublet Higgs fields, since there is no term with an odd number 
of scalar fields in the original Lagrangian. Experimentally, the triple Higgs 
($hhh$) coupling is determined from the double Higgs production processes, 
$e^+e^- \to Zhh$ and $e^+e^- \to \nu \bar{\nu} hh$ (W fusion process). The 
precision of the coupling determination  is estimated as about 20\% 
(10\%) at 500 GeV (1 TeV) LC with an integrated luminosity of 1 ab$^{-1}$.
\cite{self}

In this talk, we discuss new physics effects on the $hhh$  
coupling constant. In particular, we take a Two Higgs Doublet Model (THDM) 
as an example of physics beyond the Standard Model (SM), and consider 
relationship between the radiative correction to the $hhh$ coupling 
constant and the condition for successful electroweak baryogenesis. 
Details on the calculation are presented elsewhere.\cite{ewbg}

\section{Two Higgs Doublet Model and Electroweak Baryogenesis}
Explaining the baryon-to-photon ratio of the Universe is
a fundamental issue in the particle physics
in connection with cosmology. Although the value is $O(10^{-10})$,
it is not easy to create a sufficient baryon number from the baryon-number 
symmetric initial condition.
In 1980's, it was realized that the anomalous baryon number violation
process becomes large at high temperature in the 
SM.\cite{Kuzmin:1985mm}
This opened a possibility to generate a correct baryon number 
at the electroweak phase transition. 

In order to realize successful electroweak baryogenesis,
the electroweak phase transition has to be strong first order.
In addition, effects of CP violation should be large enough to generate
some charge flow across the expanding bubble wall that separates
the broken and unbroken phases during the phase transition. 
In the minimal SM, it is known that a sufficient
baryon number cannot be generated, 
because the phase transition is not
of the first order and the effect of CP violation due to the 
Kobayashi-Maskawa phase is far too small. Therefore the Higgs sector
should be extended for electroweak baryogenesis.

We consider the THDM as a viable model for the electroweak 
baryogenesis.\cite{thdm}
The tree-level Higgs potential of the THDM is given by
\begin{eqnarray}
 V_{tree}&=&m_1^2|\Phi_1|^2+m_2^2|\Phi_2|^2-(m_3^2\Phi_1^\dagger\Phi_2
+\mbox{h.c.})+\frac{\lambda_1}{2}|\Phi_1|^4
\nonumber\\
&&+\frac{\lambda_2}{2}|\Phi_2|^4+\lambda_3|\Phi_1|^2|\Phi_2|^2
+\lambda_4|\Phi_1^\dagger\Phi_2|^2+
\Bigg[\frac{\lambda_5}{2}(\Phi_1^\dagger\Phi_2)^2
+\mbox{h.c.}\Bigg].\label{higgs-pot}
\end{eqnarray}
In order to simplify the analysis of the electroweak phase transition,
we focus on the case with $m_1=m_2$ and $\lambda_1=\lambda_2$,
which is translated to the relation 
$\sin{(\beta-\alpha)}=\tan{\beta}=1$ for the neutral Higgs and vacuum
mixing angles. In this case, a relevant direction of the electroweak 
phase transition can be reduced to one complex dimension, and we
calculate the finite temperature effective potential for this direction
($\varphi$).

The condition for the strong first order phase transition is expressed as
\begin{equation}
\varphi_c/T_c > 1,
\label{cond}
\end{equation}
where $\varphi_c$ is the vacuum expectation value at the critical 
temperature $(T_c)$.\cite{Moore} This condition is required in order
not to erase the created baryon number by the sphaleron process 
after the phase transition.
Unlike the SM with the current Higgs-boson mass bound, 
the first order phase transition is possible in the THDM
due to a large correction to the finite temperature
effective potential from heavy Higgs boson loop diagrams. 
We can understand the qualitative feature of the effective potential
using the high temperature expansion in the case of 
$M(\equiv m_3/\sqrt{\cos{\beta}\sin{\beta}})=0$. In this case,
\begin{equation}
V_{eff}(\varphi, T)\simeq D(T^2-T^2_0)\varphi^2
-ET|\varphi|^3+\frac{\lambda_{T}}{4}\varphi^4+\cdots,
\end{equation}
where the coefficient of the cubic term is given by
$
E \simeq \frac{1}{12 \pi v^3} (6 m_W^3 + 3 m_Z^3 
 + m_H^3 + m_A^3 + 2 m_{H^\pm}^3)
$,
and $H, A,$ and $H^\pm$ are the heavy extra Higgs bosons. 
Then, $\varphi_c/T_c$ is given by $2E/\lambda_{T_c}$.
In the case of the SM, the contribution from the heavy Higgs boson
masses is missing, so that $E$ is too small for a presently allowed 
value of the Higgs boson mass.
On the other hand, the condition (\ref{cond}) can be satisfied 
for reasonable values of heavy Higgs boson masses in the THDM.

In the previous work, we have investigated radiative corrections to
the $hhh$ coupling in the THDM.\cite{Kanemura:2002vm}
We found that the correction contains quartic terms of the heavy extra 
Higgs boson masses for $M=0$ due to heavy Higgs loop diagrams. 
We therefore expect a large correction to the $hhh$ coupling 
when the above condition on the first order 
phase transition is satisfied.
  
\section{Numerical Calculation}
We evaluate the finite temperature effective 
potential and the one-loop corrected effective potential 
at zero temperature in the THDM. Since the high 
temperature expansion is not a good approximation if the heavy extra 
Higgs boson mass exceeds the critical temperature $T_c$, we calculate 
the finite temperature potential numerically. We also include the 
improvement from the ring resummation. We determine $T_c$ and the 
value of $\varphi$ at $T_c$. We also calculate the correction to 
the $hhh$ coupling constant at the zero-temperature.
(The actual calculation is done using the on-sell renormalization scheme 
instead of the effective potential method.\cite{Kanemura:2002vm}) 
In order to satisfy the $\rho$ parameter constraint on the THDM, 
we take all the heavy extra Higgs boson masses 
to be the same value $(m_{\Phi})$. 

In figure 1, we show the contour plot of the correction to the triple
Higgs coupling constant $(\Delta \lambda_{hhh}/ \lambda_{hhh})$ in the 
$(M, m_{\Phi})$ space. We overlay the line of  $\varphi_c/T_c= 1$ 
in this plot. The lightest Higgs boson mass ($m_h$) is taken to be 120 GeV.
We can see that the correction is larger than about
10\% in the case when the phase transition is strong enough for
successful electroweak baryogenesis. Such size of the deviation from the 
SM prediction is  within the reach of the measurement at a future LC.
\indent
%
     \begin{figure}[ht] 
     \begin{center}
     \vspace*{0cm}
     \epsfig{file=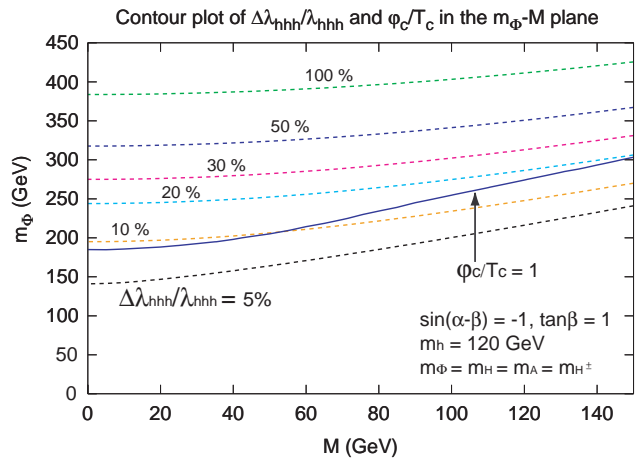,width=8.5cm}
     \end{center}
\caption{$\Delta \lambda_{hhh}/\lambda_{hhh}$ and $\varphi_c/T_c$ in THDM.
The solid line stands for the critical line which satisfy
the condition, $\varphi_c/T_c=1$. The dashed lines represent
the deviation of the $hhh$ coupling from the SM value, where
$\Delta \lambda_{hhh}\equiv \lambda_{hhh}^{eff}
(\mbox{THDM})-\lambda_{hhh}^{eff}(\mbox{SM})$.}  
\label{fig1}
     \end{figure}

In this talk, we have studied a possible impact on the LC physics 
from the scenario of electroweak baryogenesis. In order to satisfy 
the condition of the strong first-order  phase transition, the finite
temperature effective potential has to receive a large correction in the THDM.
At the same time, radiative corrections to the zero-temperature 
effective potential induce a measurable deviation in the $hhh$ coupling 
constant at a LC. For successful electroweak baryogenesis, we also need to
include CP violation effects at the phase transition. Impacts of new CP 
phases to collider physics depend on details of the electroweak baryogenesis 
scenario, and will be discussed elsewhere.

The work of YO was supported in part by a Grant-in-Aid of the Ministry 
of Education, Culture, Sports, Science, and Technology, Government of
Japan, Nos.~13640309, 13135225, and 16081211.

%

\end{document}